\documentclass[twocolumn,12pt]{iopart}

\usepackage{graphicx}

\begin{document}

\title[]{Complete analysis of hyperentangled Bell state in three degrees of freedom using Kerr effect and self-assisted mechanism}
\author{Zhi Zeng$^{1,2,3}$}
\address{{$^1$CAS Key Laboratory of Quantum Optics and Center of Cold Atom Physics, Shanghai Institute of Optics and Fine Mechanics, Chinese Academy of Sciences, Shanghai 201800, China\\}
{$^2$Wangzhijiang Innovation Center for Laser, Aerospace Laser Technology and System Department, Shanghai Institute of Optics and Fine Mechanics, Chinese Academy of Sciences, Shanghai 201800, China} \\
{$^3$Institute of Signal Processing and Transmission, Nanjing University of Posts and Telecommunications, Nanjing 210003, China}}
\eads{\mailto{zengzhiphy@yeah.net}}
\vspace{10pt}

\begin{abstract}
We present an efficient scheme for the complete hyperentangled Bell state analysis (HBSA) of photon system in polarization and two longitudinal momentum degrees of freedom (DOFs), resorting to the weak cross-Kerr nonlinearity, linear optical element and single photon detector. In the process of distinguishing the 64 hyperentangled Bell states in three DOFs, the self-assisted mechanism is embedded, which can make our scheme simple and realizable. Moreover, We have discussed the applications of this complete HBSA scheme for high-capacity quantum communication protocols that based on photonic hyperentanglement in three DOFs.
\end{abstract}

\section{Introduction}
As the counterintuitive phenomenon in physical world, quantum entanglement is the core resource that enables quantum information technologies, such as quantum teleportation \cite{QT}, quantum dense coding \cite{QDC}, entanglement swapping \cite{ES}, quantum imaging \cite{QI}, and other such tasks. Quantum system simultaneously entangled in more than one degree of freedom (DOF) is called as hyperentanglement \cite{add1}, which has been widely exploited for high-capacity quantum computation and quantum communication \cite{Du1,Du2,Du3,Du4,Du5,Du6,Du7}. One of the most important applications of hyperentanglement is that it can be used for the complete Bell state analysis (BSA) \cite{A1,A2,ZZ1}. In 1998, Kwiat \emph{et al.} theoretically showed that the complete polarization BSA can be realized with linear optics while an additional DOF is introduced \cite{A1}. In 2003, Walborn \emph{et al.} theoretically presented a simplified approach for the complete BSA with auxiliary entanglement \cite{A2}. In terms of the experimental BSA, many significant works have already been accomplished \cite{A3,A4,A5,A6,A7,A8}. For example, in 2017, Williams \emph{et al.} experimentally reported the demonstration of superdense coding over optical fiber links, taking advantage of a complete BSA enabled by hyperentanglement \cite{A5}. In 2019, Kong \emph{et al.} experimentally demonstrated the complete orbital angular momentum BSA without auxiliary photons by using linear optics and additional entanglement \cite{A8}. In many hyperentanglement-based quantum information processing protocols, the deterministic distinguishing of a set of orthogonal hyperentangled Bell state is often required for getting the information \cite{AD1,AD2,AD3}. The complete hyperentangled Bell state analysis (HBSA) for two-photon system has attracted much attention in the past decade, and research has shown that it is impossible to accomplish the complete HBSA if only linear optics is utilized and no auxiliary quantum resource is introduced \cite{2,3,4}. An efficient approach to deal with this problem is using the quantum nonlinear interactions, such as cross-Kerr nonlinearity, quantum dot spin in optical microcavity and nitrogen-vacancy center in resonator \cite{5,6,7,8,9,10,ZZ2,Z1,11,Z2,Z3,C1}. These nonlinear interactions can be used for constructing the quantum parity-check gate and quantum swap gate, both of which are useful for the complete hyperentangled state analysis. However, most of the existing complete HBSA schemes are working with photonic hyperentanglement in just two different DOFs, only a few schemes have focused on the complete HBSA for photon system in more than two DOFs \cite{12,13,14,add2,15,C2}. In 2016, Liu \emph{et al.} presented the nondestructive scheme to completely distinguish the hyperentangled Bell states in three DOFs with the help of cross-Kerr nonlinearity \cite{12}. In 2021, Zhang \emph{et al.} presented the nondestructive HBSA protocol for hyperentanglement in three DOFs, resorting to the parity-check quantum nondemolition detector \cite{14}. In 2022, Zhou \emph{et al.} proposed a simplified scheme for distinguishing the two-photon hyperentangled Bell states in three DOFs using the quantum dot-cavity interactions \cite{15}. In 2023, Sun \emph{et al.} presented a complete and fidelity-robust HBSA scheme for two-photon polarization-spatial-time-bin hyperentanglement with double-sided quantum-dot-cavity systems \cite{C2}.

Compared with the traditional hyperentangled state in two DOFs, hyperentanglement in three DOFs can largely increase the channel capacity of quantum communication, which has already been efficiently prepared and manipulated in experiment \cite{16,17,add3,add4}. In 2009, a six-qubit hyperentangled Bell state was realized by entangling two photons in three different DOFs, including the polarization DOF and two longitudinal momentum DOFs \cite{17}. By using this experimentally available hyperentanglement, Li \emph{et al.} presented a hyperdense coding scheme based on their novel linear optical HBSA technique in 2017 \cite{3}. They found that the 16 hyperentangled Bell states in two DOFs can be classified into 12 groups via just linear optics, resorting to a fixed auxiliary Bell state in the third DOF \cite{3}. With $n$ DOFs, the $4^n$ hyperentangled Bell states can be separated into $x_k = 2^{n+k+1} - 2^{2k}$ groups via linear optics, resorting to the $k(k\leq n)$ ancillary entangled states \cite{3}. When $k = n$, all the $4^n$ hyperentangled Bell states can be completely discriminated. Inspired by this significant work, in this paper, we present a simple and efficient scheme for the complete analysis of hyperentangled Bell state in polarization and two longitudinal momentum DOFs, by using the cross-Kerr nonlinearity, linear optical element and single photon detector. In our complete HBSA scheme for two-photon system in three DOFs, all the 64 hyperentangled Bell states can be distinguished. With the help of the weak cross-Kerr nonlinearity and self-assisted mechanism, the discrimination process is greatly simplified and the whole scheme becomes more realizable. Our complete HBSA scheme can be directly used for the quantum teleportation and entanglement swapping of photon system in three DOFs, and will have useful applications in other high-capacity quantum information processing tasks.

\section{Complete HBSA for hyperentanglement in three degrees of freedom}
The general form of two-photon hyperentangled Bell state in polarization and two longitudinal momentum DOFs can be written as
\begin{eqnarray}
|\Psi\rangle_{AB} = |\Psi\rangle_{P} \otimes |\Psi\rangle_{F} \otimes |\Psi\rangle_{S}.
\end{eqnarray}
Here, $A$ and $B$ are the two entangled photons. $P$, $F$ and $S$ denote the polarization DOF, the first longitudinal momentum DOF and the second longitudinal momentum DOF, respectively. $|\Psi\rangle_{P}$ is one of the four Bell states in polarization DOF,
\begin{eqnarray}
|\phi^{\pm}\rangle_{P} = \frac{1}{\sqrt 2} (|HH\rangle \pm |VV\rangle)_{AB}, \nonumber \\
|\psi^{\pm}\rangle_{P} = \frac{1}{\sqrt 2} (|HV\rangle \pm |VH\rangle)_{AB},
\end{eqnarray}
where $|H\rangle$ and $|V\rangle$ are the horizontal and vertical polarization states of photon, respectively. $|\Psi\rangle_{F}$ is one of the four Bell states in the first longitudinal momentum DOF,
\begin{eqnarray}
|\phi^{\pm}\rangle_{F} = \frac{1}{\sqrt 2} (|EE\rangle \pm |II\rangle)_{AB}, \nonumber \\
|\psi^{\pm}\rangle_{F} = \frac{1}{\sqrt 2} (|EI\rangle \pm |IE\rangle)_{AB},
\end{eqnarray}
where $|E\rangle$ and $|I\rangle$ are the external and internal spatial-modes of photon, respectively. $|\Psi\rangle_{S}$ is one of the four Bell states in the second longitudinal momentum DOF,
\begin{eqnarray}
|\phi^{\pm}\rangle_{S} = \frac{1}{\sqrt 2} (|rr\rangle \pm |ll\rangle)_{AB}, \nonumber \\
|\psi^{\pm}\rangle_{S} = \frac{1}{\sqrt 2} (|rl\rangle \pm |lr\rangle)_{AB},
\end{eqnarray}
where $|r\rangle$ and $|l\rangle$ are the right and left spatial-modes of photon, respectively. Considering the three DOFs at the same time, there are 64 orthogonal hyperentangled Bell states, which will be unambiguously distinguished in the following text.

Before describing our complete HBSA scheme, we first briefly introduce the principle of nondestructive detection of photon number, which is achieved by using the weak cross-Kerr nonlinearity. The interaction between a signal state $|\varphi\rangle_s$ and a probe coherent state $|\alpha\rangle_p$ in a nonlinear medium can be described with the Hamiltonian \cite{Kerr1}
\begin{eqnarray}
H = \hbar\chi a^\dag_sa_sa^\dag_pa_p.
\end{eqnarray}
Here, $a^\dag_s$ ($a^\dag_p$) and $a_s$ ($a_p$) are the creation and destruction operators of the signal (probe) state, respectively. $\chi$ is the coupling strength of nonlinearity, which depends on the property of nonlinear material. After the interaction with the signal state in nonlinear medium, the probe coherent state can get a phase shift, which is proportional to the photon number $N$ of the signal state,
\begin{eqnarray}
|\alpha\rangle_p \rightarrow |\alpha e^{iN\theta}\rangle.
\end{eqnarray}
Here $\theta = \chi t$ ($t$ is the interaction time). When we use the $X$-quadrature measurement on the phase shift of probe coherent state, the photon number of signal state can be read out without destroying the photons.

\begin{figure}
\centering
\includegraphics*[width=0.9\textwidth]{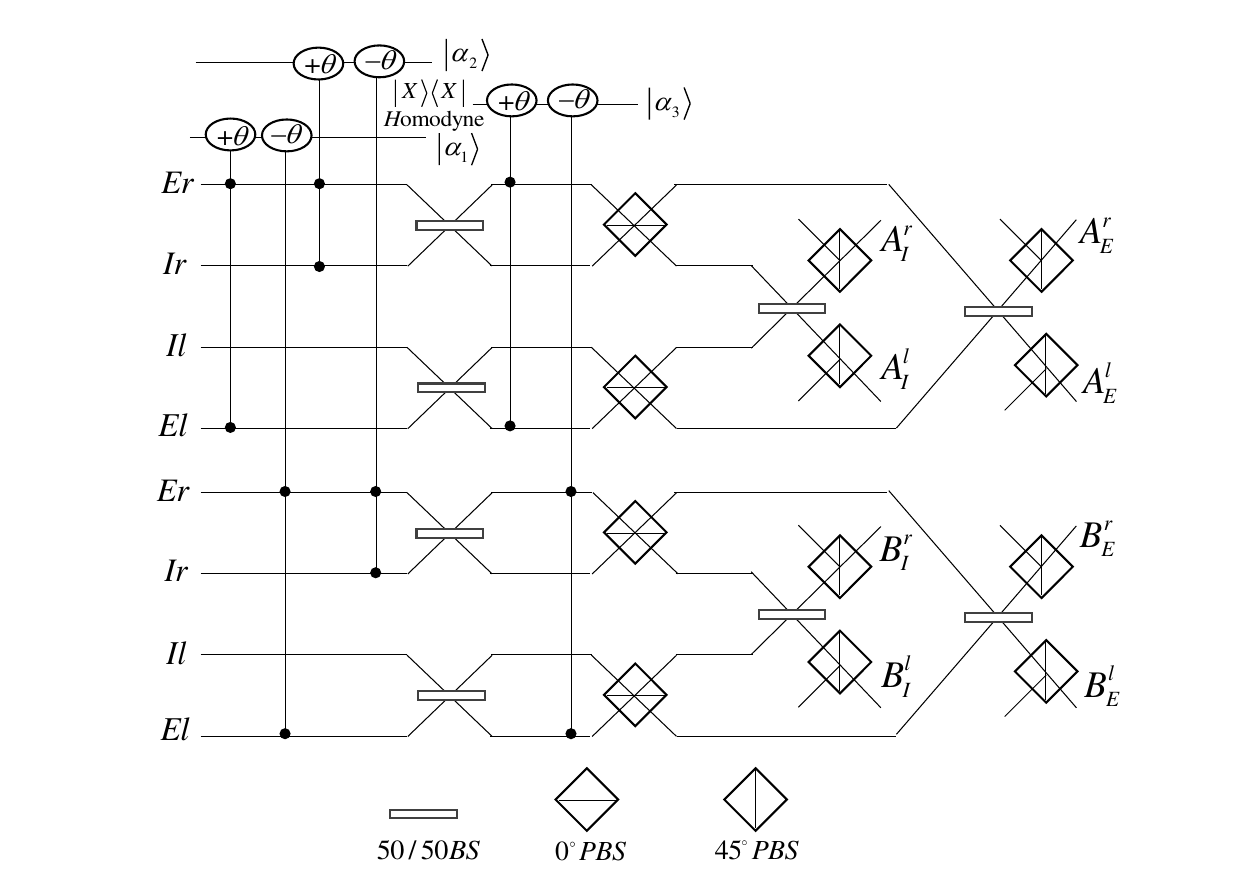}
\caption{Schematic diagram of our complete HBSA scheme for photon system in polarization and two longitudinal momentum DOFs. The cross-Kerr nonlinear interaction can generate the phase shift of $\pm\theta$ on coherent state, when photon appears in the corresponding spatial-mode. The 50:50 beam splitter (BS) performs the Hadamard operation [$|x_1\rangle\rightarrow(|x_1\rangle + |x_2\rangle)/{\sqrt 2}, |x_2\rangle\rightarrow(|x_1\rangle - |x_2\rangle)/{\sqrt 2}$] on the spatial-mode state of photon. The polarization beam splitter (PBS) orientated at $0^{\circ}$ transmits the $|H\rangle$ photon and reflects the $|V\rangle$ photon, while the PBS orientated at $45^{\circ}$ transmits the $|+\rangle=(|H\rangle + |V\rangle)/{\sqrt 2}$ photon and reflects the $|-\rangle=(|H\rangle - |V\rangle)/{\sqrt 2}$ photon. Single photon detector is required to accomplish the polarization orthogonal measurement in \{$|+\rangle$, $|-\rangle$\} basis. With this setup, the 64 hyperentangled Bell states in three DOFs can be completely discriminated from each other.}
\end{figure}

The setup of our complete HBSA scheme for three-DOF hyperentanglement is shown in Fig. 1. After the two photons $A$ and $B$ interact with the coherent states $|\alpha_1\rangle$ and $|\alpha_2\rangle$, the state of the collective system evolves as
\begin{eqnarray}
|\Psi\rangle_{P}|\phi^{\pm}\rangle_{F}|\phi^{\pm}\rangle_{S}|\alpha_1\rangle|\alpha_2\rangle &\rightarrow&  |\Psi\rangle_{P}|\phi^{\pm}\rangle_{F}|\phi^{\pm}\rangle_{S}|\alpha_1\rangle|\alpha_2\rangle,   \nonumber \\
|\Psi\rangle_{P}|\phi^{\pm}\rangle_{F}|\psi^{\pm}\rangle_{S}|\alpha_1\rangle|\alpha_2\rangle &\rightarrow&  |\Psi\rangle_{P}|\phi^{\pm}\rangle_{F}|\psi^{\pm}\rangle_{S}|\alpha_1\rangle|\alpha_2e^{\pm i\theta}\rangle,   \nonumber \\
|\Psi\rangle_{P}|\psi^{\pm}\rangle_{F}|\phi^{\pm}\rangle_{S}|\alpha_1\rangle|\alpha_2\rangle &\rightarrow&  |\Psi\rangle_{P}|\psi^{\pm}\rangle_{F}|\phi^{\pm}\rangle_{S}|\alpha_1e^{\pm i\theta}\rangle|\alpha_2\rangle,   \nonumber \\
|\Psi\rangle_{P}|\psi^{\pm}\rangle_{F}|\psi^{\pm}\rangle_{S}|\alpha_1\rangle|\alpha_2\rangle &\rightarrow&  |\Psi\rangle_{P}|\psi^{\pm}\rangle_{F}|\psi^{\pm}\rangle_{S}|\alpha_1e^{\pm i\theta}\rangle|\alpha_2e^{\pm i\theta}\rangle.
\end{eqnarray}
The polarization entanglement is invariant during the interaction, and the parity information of entanglement in the two longitudinal momentum DOFs can be determined through the two coherent states. Then, the two photons will pass through BSs, interact with coherent state $|\alpha_3\rangle$, and the state of the collective system evolves as
\begin{eqnarray}
|\Psi\rangle_{P}|\phi^{+}\rangle_{F}|\Psi\rangle_{S}|\alpha_3\rangle &\rightarrow& |\Psi\rangle_{P}|\phi^{+}\rangle_{F}|\Psi\rangle_{S}|\alpha_3\rangle,   \nonumber \\
|\Psi\rangle_{P}|\phi^{-}\rangle_{F}|\Psi\rangle_{S}|\alpha_3\rangle &\rightarrow& |\Psi\rangle_{P}|\psi^{+}\rangle_{F}|\Psi\rangle_{S}|\alpha_3e^{\pm i\theta}\rangle,   \nonumber \\
|\Psi\rangle_{P}|\psi^{+}\rangle_{F}|\Psi\rangle_{S}|\alpha_3\rangle &\rightarrow& |\Psi\rangle_{P}|\phi^{-}\rangle_{F}|\Psi\rangle_{S}|\alpha_3\rangle,   \nonumber \\
|\Psi\rangle_{P}|\psi^{-}\rangle_{F}|\Psi\rangle_{S}|\alpha_3\rangle &\rightarrow& |\Psi\rangle_{P}|\psi^{-}\rangle_{F}|\Psi\rangle_{S}|\alpha_3e^{\pm i\theta}\rangle.
\end{eqnarray}
We can find that the entanglement in polarization and the second longitudinal momentum DOFs are both unchanged, and the phase information of entanglement in the first longitudinal momentum DOF can be determined through coherent state $|\alpha_3\rangle$. With the help of measurements on the three coherent states, the 64 hyperentangled Bell states in three DOFs can be divided into eight groups, as shown in Table 1.

\begin{table}
\centering\caption{Relations between the initial states and phase shifts of the three coherent states.}
\begin{tabular}{cc ccccccccccc}
\hline
Initial states & & & & $|\alpha_1\rangle$ & & & & $|\alpha_2\rangle$ & & & & $|\alpha_3\rangle$ \\
\hline
$|\Psi\rangle_{P}|\phi^{+}\rangle_{F}|\phi^{\pm}\rangle_{S}$ &&&& 0 &&&& 0 &&&& 0 \\
$|\Psi\rangle_{P}|\phi^{-}\rangle_{F}|\phi^{\pm}\rangle_{S}$ &&&& 0 &&&& 0 &&&& $\pm\theta$ \\
$|\Psi\rangle_{P}|\phi^{+}\rangle_{F}|\psi^{\pm}\rangle_{S}$ &&&& 0 &&&& $\pm\theta$ &&&& 0 \\
$|\Psi\rangle_{P}|\phi^{-}\rangle_{F}|\psi^{\pm}\rangle_{S}$ &&&& 0 &&&& $\pm\theta$ &&&& $\pm\theta$ \\
$|\Psi\rangle_{P}|\psi^{+}\rangle_{F}|\phi^{\pm}\rangle_{S}$ &&&& $\pm\theta$ &&&& 0 &&&& 0 \\
$|\Psi\rangle_{P}|\psi^{-}\rangle_{F}|\phi^{\pm}\rangle_{S}$ &&&& $\pm\theta$ &&&& 0 &&&& $\pm\theta$ \\
$|\Psi\rangle_{P}|\psi^{+}\rangle_{F}|\psi^{\pm}\rangle_{S}$ &&&& $\pm\theta$ &&&& $\pm\theta$ &&&& 0 \\
$|\Psi\rangle_{P}|\psi^{-}\rangle_{F}|\psi^{\pm}\rangle_{S}$ &&&& $\pm\theta$ &&&& $\pm\theta$ &&&& $\pm\theta$ \\
\hline
\end{tabular}
\end{table}

There are eight hyperentangled Bell states in each group, which can be distinguished by using linear optical element and single photon detector. Here, We take the first group in Table 1 as an example to illustrate. After passing through BSs, PBSs and BSs, the evolution of these initial eight states can be expressed as
\begin{eqnarray}
|\phi^{+}\rangle_{P}|\phi^{+}\rangle_{F}|\phi^{+}\rangle_{S} &\rightarrow&  \frac{1}{2\sqrt 2}(|++\rangle + |--\rangle)(|EE\rangle + |II\rangle)(|rr\rangle + |ll\rangle)_{AB},   \nonumber \\
|\phi^{-}\rangle_{P}|\phi^{+}\rangle_{F}|\phi^{+}\rangle_{S} &\rightarrow&  \frac{1}{2\sqrt 2}(|+-\rangle + |-+\rangle)(|EE\rangle + |II\rangle)(|rr\rangle + |ll\rangle)_{AB},   \nonumber \\
|\psi^{+}\rangle_{P}|\phi^{+}\rangle_{F}|\phi^{+}\rangle_{S} &\rightarrow&  \frac{1}{2\sqrt 2}(|++\rangle - |--\rangle)(|EI\rangle + |IE\rangle)(|rr\rangle + |ll\rangle)_{AB},   \nonumber \\
|\psi^{-}\rangle_{P}|\phi^{+}\rangle_{F}|\phi^{+}\rangle_{S} &\rightarrow&  \frac{1}{2\sqrt 2}(|+-\rangle - |-+\rangle)(|EI\rangle + |IE\rangle)(|rr\rangle + |ll\rangle)_{AB},   \nonumber \\
|\phi^{+}\rangle_{P}|\phi^{+}\rangle_{F}|\phi^{-}\rangle_{S} &\rightarrow&  \frac{1}{2\sqrt 2}(|++\rangle + |--\rangle)(|EE\rangle + |II\rangle)(|rl\rangle + |lr\rangle)_{AB},   \nonumber \\
|\phi^{-}\rangle_{P}|\phi^{+}\rangle_{F}|\phi^{-}\rangle_{S} &\rightarrow&  \frac{1}{2\sqrt 2}(|+-\rangle + |-+\rangle)(|EE\rangle + |II\rangle)(|rl\rangle + |lr\rangle)_{AB},   \nonumber \\
|\psi^{+}\rangle_{P}|\phi^{+}\rangle_{F}|\phi^{-}\rangle_{S} &\rightarrow&  \frac{1}{2\sqrt 2}(|++\rangle - |--\rangle)(|EI\rangle + |IE\rangle)(|rl\rangle + |lr\rangle)_{AB},   \nonumber \\
|\psi^{-}\rangle_{P}|\phi^{+}\rangle_{F}|\phi^{-}\rangle_{S} &\rightarrow&  \frac{1}{2\sqrt 2}(|+-\rangle - |-+\rangle)(|EI\rangle + |IE\rangle)(|rl\rangle + |lr\rangle)_{AB}. 
\end{eqnarray}
The above eight hyperentangled Bell states in three DOFs can be identified by using the detection results of single photon detectors, with which the initial 64 states can be classified into eight groups, as shown in Table 2. Based on both Table 1 and Table 2, the 64 hyperentangled Bell states can be completely distinguished. 

\begin{table}[!htbp]
\centering\caption{The initial 64 hyperentangled Bell states in three DOFs can be classified into eight groups with the detection results of single photon detectors.}
\begin{tabular}{cc ccccccccccc ccccccccccc}
\hline
Group &  Initial states & Detection results  \\
\hline
$1$ & $|\phi^{+}\rangle_{P}|\phi^{+}\rangle_{F}|\phi^{+}\rangle_{S},|\phi^{+}\rangle_{P}|\psi^{+}\rangle_{F}|\phi^{+}\rangle_{S},$ & $A_{E}^{r+}B_{E}^{r+},A_{E}^{r-}B_{E}^{r-},$  \\ &
$|\psi^{+}\rangle_{P}|\phi^{-}\rangle_{F}|\phi^{+}\rangle_{S},|\psi^{+}\rangle_{P}|\psi^{-}\rangle_{F}|\phi^{+}\rangle_{S},$ & $A_{E}^{l+}B_{E}^{l+},A_{E}^{l-}B_{E}^{l-},$   \\ &
$|\phi^{+}\rangle_{P}|\phi^{+}\rangle_{F}|\psi^{+}\rangle_{S},|\phi^{+}\rangle_{P}|\psi^{+}\rangle_{F}|\psi^{+}\rangle_{S},$ & $A_{I}^{r+}B_{I}^{r+},A_{I}^{r-}B_{I}^{r-},$   \\ &
$|\psi^{+}\rangle_{P}|\phi^{-}\rangle_{F}|\psi^{+}\rangle_{S},|\psi^{+}\rangle_{P}|\psi^{-}\rangle_{F}|\psi^{+}\rangle_{S}.$ & $A_{I}^{l+}B_{I}^{l+},A_{I}^{l-}B_{I}^{l-}.$   \\

$2$ & $|\phi^{+}\rangle_{P}|\phi^{-}\rangle_{F}|\phi^{+}\rangle_{S},|\phi^{+}\rangle_{P}|\psi^{-}\rangle_{F}|\phi^{+}\rangle_{S},$ & $A_{E}^{r+}B_{I}^{r+},A_{E}^{r-}B_{I}^{r-},$  \\ &
$|\psi^{+}\rangle_{P}|\phi^{+}\rangle_{F}|\phi^{+}\rangle_{S},|\psi^{+}\rangle_{P}|\psi^{+}\rangle_{F}|\phi^{+}\rangle_{S},$ & $A_{E}^{l+}B_{I}^{l+},A_{E}^{l-}B_{I}^{l-},$  \\ &
$|\phi^{+}\rangle_{P}|\phi^{-}\rangle_{F}|\psi^{+}\rangle_{S},|\phi^{+}\rangle_{P}|\psi^{-}\rangle_{F}|\psi^{+}\rangle_{S},$ & $A_{I}^{r+}B_{E}^{r+},A_{I}^{r-}B_{E}^{r-},$  \\ &
$|\psi^{+}\rangle_{P}|\phi^{+}\rangle_{F}|\psi^{+}\rangle_{S},|\psi^{+}\rangle_{P}|\psi^{+}\rangle_{F}|\psi^{+}\rangle_{S}.$ & $A_{I}^{l+}B_{E}^{l+},A_{I}^{l-}B_{E}^{l-}.$   \\

$3$ & $|\phi^{-}\rangle_{P}|\phi^{+}\rangle_{F}|\phi^{+}\rangle_{S},|\phi^{-}\rangle_{P}|\psi^{+}\rangle_{F}|\phi^{+}\rangle_{S},$ & $A_{E}^{r+}B_{E}^{r-},A_{E}^{r-}B_{E}^{r+},$  \\ &
$|\psi^{-}\rangle_{P}|\phi^{-}\rangle_{F}|\phi^{+}\rangle_{S},|\psi^{-}\rangle_{P}|\psi^{-}\rangle_{F}|\phi^{+}\rangle_{S},$ & $A_{E}^{l+}B_{E}^{l-},A_{E}^{l-}B_{E}^{l+},$  \\ &
$|\phi^{-}\rangle_{P}|\phi^{+}\rangle_{F}|\psi^{+}\rangle_{S},|\phi^{-}\rangle_{P}|\psi^{+}\rangle_{F}|\psi^{+}\rangle_{S},$ & $A_{I}^{r+}B_{I}^{r-},A_{I}^{r-}B_{I}^{r+},$   \\ &
$|\psi^{-}\rangle_{P}|\phi^{-}\rangle_{F}|\psi^{+}\rangle_{S},|\psi^{-}\rangle_{P}|\psi^{-}\rangle_{F}|\psi^{+}\rangle_{S}.$ & $A_{I}^{l+}B_{I}^{l-},A_{I}^{l-}B_{I}^{l+}.$    \\

$4$ & $|\phi^{-}\rangle_{P}|\phi^{-}\rangle_{F}|\phi^{+}\rangle_{S},|\phi^{-}\rangle_{P}|\psi^{-}\rangle_{F}|\phi^{+}\rangle_{S},$ & $A_{E}^{r+}B_{I}^{r-},A_{E}^{r-}B_{I}^{r+},$  \\ &
$|\psi^{-}\rangle_{P}|\phi^{+}\rangle_{F}|\phi^{+}\rangle_{S},|\psi^{-}\rangle_{P}|\psi^{+}\rangle_{F}|\phi^{+}\rangle_{S},$ & $A_{E}^{l+}B_{I}^{l-},A_{E}^{l-}B_{I}^{l+},$   \\ &
$|\phi^{-}\rangle_{P}|\phi^{-}\rangle_{F}|\psi^{+}\rangle_{S},|\phi^{-}\rangle_{P}|\psi^{-}\rangle_{F}|\psi^{+}\rangle_{S},$ & $A_{I}^{r+}B_{E}^{r-},A_{I}^{r-}B_{E}^{r+},$   \\ &
$|\psi^{-}\rangle_{P}|\phi^{+}\rangle_{F}|\psi^{+}\rangle_{S},|\psi^{-}\rangle_{P}|\psi^{+}\rangle_{F}|\psi^{+}\rangle_{S}.$ & $A_{I}^{l+}B_{E}^{l-},A_{I}^{l-}B_{E}^{l+}.$  \\

$5$ & $|\phi^{+}\rangle_{P}|\phi^{+}\rangle_{F}|\phi^{-}\rangle_{S},|\phi^{+}\rangle_{P}|\psi^{+}\rangle_{F}|\phi^{-}\rangle_{S},$ & $A_{E}^{r+}B_{E}^{l+},A_{E}^{r-}B_{E}^{l-},$   \\ &
$|\psi^{+}\rangle_{P}|\phi^{-}\rangle_{F}|\phi^{-}\rangle_{S},|\psi^{+}\rangle_{P}|\psi^{-}\rangle_{F}|\phi^{-}\rangle_{S},$ & $A_{E}^{l+}B_{E}^{r+},A_{E}^{l-}B_{E}^{r-},$   \\ &
$|\phi^{+}\rangle_{P}|\phi^{+}\rangle_{F}|\psi^{-}\rangle_{S},|\phi^{+}\rangle_{P}|\psi^{+}\rangle_{F}|\psi^{-}\rangle_{S},$ & $A_{I}^{r+}B_{I}^{l+},A_{I}^{r-}B_{I}^{l-},$    \\ &
$|\psi^{+}\rangle_{P}|\phi^{-}\rangle_{F}|\psi^{-}\rangle_{S},|\psi^{+}\rangle_{P}|\psi^{-}\rangle_{F}|\psi^{-}\rangle_{S}.$ & $A_{I}^{l+}B_{I}^{r+},A_{I}^{l-}B_{I}^{r-}.$   \\

$6$ & $|\phi^{+}\rangle_{P}|\phi^{-}\rangle_{F}|\phi^{-}\rangle_{S},|\phi^{+}\rangle_{P}|\psi^{-}\rangle_{F}|\phi^{-}\rangle_{S},$ & $A_{E}^{r+}B_{I}^{l+},A_{E}^{r-}B_{I}^{l-},$  \\ &
$|\psi^{+}\rangle_{P}|\phi^{+}\rangle_{F}|\phi^{-}\rangle_{S},|\psi^{+}\rangle_{P}|\psi^{+}\rangle_{F}|\phi^{-}\rangle_{S},$ & $A_{E}^{l+}B_{I}^{r+},A_{E}^{l-}B_{I}^{r-},$   \\ &
$|\phi^{+}\rangle_{P}|\phi^{-}\rangle_{F}|\psi^{-}\rangle_{S},|\phi^{+}\rangle_{P}|\psi^{-}\rangle_{F}|\psi^{-}\rangle_{S},$ & $A_{I}^{r+}B_{E}^{l+},A_{I}^{r-}B_{E}^{l-},$    \\ &
$|\psi^{+}\rangle_{P}|\phi^{+}\rangle_{F}|\psi^{-}\rangle_{S},|\psi^{+}\rangle_{P}|\psi^{+}\rangle_{F}|\psi^{-}\rangle_{S}.$ & $A_{I}^{l+}B_{E}^{r+},A_{I}^{l-}B_{E}^{r-}.$ \\

$7$ & $|\phi^{-}\rangle_{P}|\phi^{+}\rangle_{F}|\phi^{-}\rangle_{S},|\phi^{-}\rangle_{P}|\psi^{+}\rangle_{F}|\phi^{-}\rangle_{S},$  & $A_{E}^{r+}B_{E}^{l-},A_{E}^{r-}B_{E}^{l+},$   \\ &
$|\psi^{-}\rangle_{P}|\phi^{-}\rangle_{F}|\phi^{-}\rangle_{S},|\psi^{-}\rangle_{P}|\psi^{-}\rangle_{F}|\phi^{-}\rangle_{S},$ & $A_{E}^{l+}B_{E}^{r-},A_{E}^{l-}B_{E}^{r+},$   \\ &
$|\phi^{-}\rangle_{P}|\phi^{+}\rangle_{F}|\psi^{-}\rangle_{S},|\phi^{-}\rangle_{P}|\psi^{+}\rangle_{F}|\psi^{-}\rangle_{S},$ & $A_{I}^{r+}B_{I}^{l-},A_{I}^{r-}B_{I}^{l+},$  \\ &
$|\psi^{-}\rangle_{P}|\phi^{-}\rangle_{F}|\psi^{-}\rangle_{S},|\psi^{-}\rangle_{P}|\psi^{-}\rangle_{F}|\psi^{-}\rangle_{S}.$ & $A_{I}^{l+}B_{I}^{r-},A_{I}^{l-}B_{I}^{r+}.$   \\

$8$ & $|\phi^{-}\rangle_{P}|\phi^{-}\rangle_{F}|\phi^{-}\rangle_{S},|\phi^{-}\rangle_{P}|\psi^{-}\rangle_{F}|\phi^{-}\rangle_{S},$ & $A_{E}^{r+}B_{I}^{l-},A_{E}^{r-}B_{I}^{l+},$   \\ &
$|\psi^{-}\rangle_{P}|\phi^{+}\rangle_{F}|\phi^{-}\rangle_{S},|\psi^{-}\rangle_{P}|\psi^{+}\rangle_{F}|\phi^{-}\rangle_{S},$ & $A_{E}^{l+}B_{I}^{r-},A_{E}^{l-}B_{I}^{r+},$    \\ &
$|\phi^{-}\rangle_{P}|\phi^{-}\rangle_{F}|\psi^{-}\rangle_{S},|\phi^{-}\rangle_{P}|\psi^{-}\rangle_{F}|\psi^{-}\rangle_{S},$ & $A_{I}^{r+}B_{E}^{l-},A_{I}^{r-}B_{E}^{l+},$    \\ &
$|\psi^{-}\rangle_{P}|\phi^{+}\rangle_{F}|\psi^{-}\rangle_{S},|\psi^{-}\rangle_{P}|\psi^{+}\rangle_{F}|\psi^{-}\rangle_{S}.$ & $A_{I}^{l+}B_{E}^{r-},A_{I}^{l-}B_{E}^{r+}.$    \\
\hline
\end{tabular}
\end{table}

Owing to the Kerr effect and self-assisted mechanism, our complete HBSA scheme is simple and efficient. Specifically speaking, the four Bell states in the first longitudinal momentum DOF and the parity information of entanglement in the second longitudinal momentum DOF are determined by using the weak cross-Kerr nonlinearity. After that, the polarization Bell states are distinguished through the preserved entanglement in the first longitudinal momentum DOF, and the phase information of Bell states in the second longitudinal momentum DOF are identified by using linear optical element and single photon detector.

\section{Applications of our complete HBSA scheme in quantum communication}
Hyperentangled state analysis is necessary for many high-capacity quantum communication protocols, such as hyperentanglement-based quantum teleportation, dense coding, entanglement swapping, quantum key distribution, quantum repeater and so on. In this section, we just show the applications of our complete HBSA scheme in the quantum teleportation and entanglement swapping protocols that based on the hyperentangled Bell state in polarization and two longitudinal momentum DOFs. 

\subsection{Quantum teleportation of a single-photon state in three DOFs}
Suppose the sender Alice wants to send the receiver Bob an unknown quantum state of photon $X$, which is a single-photon state in three DOFs,
\begin{eqnarray}
|\phi\rangle_{X} = (\alpha_P|H\rangle + \beta_P|V\rangle) \otimes (\alpha_F|E\rangle + \beta_F|I\rangle) \otimes (\alpha_S|r\rangle + \beta_S|l\rangle)_X.
\end{eqnarray}
The two communication parties have shared a hyperentangled Bell state in three DOFs as the quantum channel in advance
\begin{eqnarray}
|\Phi\rangle_{AB} = \frac{1}{2\sqrt 2}(|HH\rangle + |VV\rangle)\otimes(|EE\rangle + |II\rangle)\otimes(|rr\rangle+|ll\rangle)_{AB}.
\end{eqnarray}
After Alice performs HBSA on the two photons $X$ and $A$, the quantum state of the three-photon system can be rewritten as the following form,
\begin{eqnarray}
|\phi\rangle_{X}\otimes|\Phi\rangle_{AB} = &\frac{1}{8}[|\phi^{+}\rangle_{P}(\alpha_P|H\rangle + \beta_P|V\rangle)_X +|\phi^{-}\rangle_{P}(\alpha_P|H\rangle - \beta_P|V\rangle)_X \nonumber \\ & + |\psi^{+}\rangle_{P}(\alpha_P|V\rangle + \beta_P|H\rangle)_X +|\psi^{-}\rangle_{P}(\alpha_P|V\rangle - \beta_P|H\rangle)_X] \nonumber \\ & \otimes [|\phi^{+}\rangle_{F}(\alpha_F|E\rangle + \beta_F|I\rangle)_X +|\phi^{-}\rangle_{F}(\alpha_F|E\rangle - \beta_F|I\rangle)_X \nonumber \\ & + |\psi^{+}\rangle_{F}(\alpha_F|I\rangle + \beta_F|E\rangle)_X +|\psi^{-}\rangle_{F}(\alpha_F|I\rangle - \beta_F|E\rangle)_X] \nonumber \\ & \otimes [|\phi^{+}\rangle_{S}(\alpha_S|r\rangle + \beta_S|l\rangle)_X +|\phi^{-}\rangle_{S}(\alpha_S|r\rangle - \beta_S|l\rangle)_X \nonumber \\ & + |\psi^{+}\rangle_{S}(\alpha_S|l\rangle + \beta_S|r\rangle)_X +|\psi^{-}\rangle_{S}(\alpha_S|l\rangle - \beta_S|r\rangle)_X].
\end{eqnarray}

We can find that the sender Alice has 64 possible measurement results, corresponding to which there are 64 potential single-photon states in three DOFs for the photon of receiver Bob. With the help of our complete HBSA scheme, the 64 hyperentangled Bell states in three DOFs can be unambiguously discriminated, corresponding to which Bob can get the information of the quantum state of his own photon. Based on the measurement result of Alice, Bob can map the original state of photon $X$ to his photon with proper single-photon unitary operations on the three different DOFs.

\subsection{Entanglement swapping between hyperentangled pairs in three DOFs}
Entanglement swapping enables the two parties in quantum communication to establish quantum entanglement with each other without any direct interactions between them. Suppose two remote parties Alice and Bob each share the three-DOF hyperentangled Bell state with a central node Charlie.
\begin{eqnarray}
|\Phi\rangle_{AC} = \frac{1}{2\sqrt 2}(|HH\rangle + |VV\rangle)\otimes(|EE\rangle + |II\rangle)\otimes(|rr\rangle+|ll\rangle)_{AC},  \nonumber \\
|\Phi\rangle_{BD} = \frac{1}{2\sqrt 2}(|HH\rangle + |VV\rangle)\otimes(|EE\rangle + |II\rangle)\otimes(|rr\rangle+|ll\rangle)_{BD}.
\end{eqnarray}

Here photons $A$ and $B$ belong to Alice and Bob, respectively, and photons $C$ and $D$ are held by Charlie. After Charlie performs the complete HBSA on his photons $C$ and $D$, Alice and Bob's photons will collapse into a hyperentangled Bell state.
\begin{eqnarray}
|\Phi\rangle_{AC}\otimes|\Phi\rangle_{BD} = &\frac{1}{8}(|\phi^{+}\rangle_{P}^{AB}|\phi^{+}\rangle_{P}^{CD} +|\phi^{-}\rangle_{P}^{AB}|\phi^{-}\rangle_{P}^{CD} + |\psi^{+}\rangle_{P}^{AB}|\psi^{+}\rangle_{P}^{CD} +|\psi^{-}\rangle_{P}^{AB}|\psi^{-}\rangle_{P}^{CD}) \nonumber \\ &\otimes (|\phi^{+}\rangle_{F}^{AB}|\phi^{+}\rangle_{F}^{CD} +|\phi^{-}\rangle_{F}^{AB}|\phi^{-}\rangle_{F}^{CD} + |\psi^{+}\rangle_{F}^{AB}|\psi^{+}\rangle_{F}^{CD} +|\psi^{-}\rangle_{F}^{AB}|\psi^{-}\rangle_{F}^{CD}) \nonumber \\ &\otimes (|\phi^{+}\rangle_{S}^{AB}|\phi^{+}\rangle_{S}^{CD} +|\phi^{-}\rangle_{S}^{AB}|\phi^{-}\rangle_{S}^{CD} + |\psi^{+}\rangle_{S}^{AB}|\psi^{+}\rangle_{S}^{CD} +|\psi^{-}\rangle_{S}^{AB}|\psi^{-}\rangle_{S}^{CD}). \nonumber \\
\end{eqnarray}

From the above expression, we can find that the quantum state of $AB$ depends on the measurement result of $CD$. For instance, if Charlie's result is $|\phi^{+}\rangle_{P}^{CD}|\psi^{-}\rangle_{F}^{CD}|\phi^{-}\rangle_{S}^{CD}$, the hyperentangled state shared by Alice and Bob is $|\phi^{+}\rangle_{P}^{AB}|\psi^{-}\rangle_{F}^{AB}|\phi^{-}\rangle_{S}^{AB}$. Based on Charlie's measurement result, Alice and Bob can share the desired hyperentangled Bell state with or without additional single-photon operation. With the help of our complete HBSA scheme, three-DOF hyperentanglement can be established between distant parties, which can be useful for the long-distance quantum communication.

\section{Discussion and summary}
In our complete HBSA scheme, the weak cross-Kerr nonlinearity is utilized for constructing the photon number quantum nondemolition detector (QND), the efficiency of which will directly influence the overall success probability of the whole scheme. Although many works have been reported on the cross-Kerr effect, we should acknowledge that a clean cross-Kerr nonlinearity is still challenging in the single-photon regime with the current technology \cite{Kerr2,Kerr3,Kerr4}. For a weak cross-Kerr nonlinearity, it is possible for us to distinguish the small phase shift in coherent state from the zero phase shift, when a sufficiently large amplitude of coherent state satisfies $\alpha\theta^2\gg1$ ($\theta$ is the cross-phase shift). Recent studies have shown that it is promising for us to utilize the cross-Kerr nonlinearity in the near future \cite{Kerr5,Kerr6,Kerr7,Kerr8,Kerr9,Kerr10,Kerr11,Kerr12,Kerr13}. For example, in 2009, Matsuda \emph{et al.} presented the first experimental demonstration of single-photon-level nonlinear phase shift in an optical fibre \cite{Kerr5}. In 2016, Beck \emph{et al.} measured a conditional cross-phase shift of $\pi/6$ between the retrieved signal and control photons \cite{Kerr8}. In the same year, Tiarks \emph{et al.} experimentally demonstrate the generation of the $\pi$ phase shift with a single-photon pulse \cite{Kerr9}. In 2019, Sinclair \emph{et al.} reported the experimental observation of a cross-Kerr nonlinearity in a free-space medium, which is used to implement cross-phase modulation between two optical pulses \cite{Kerr10}. Actually, we just need the small phase shift that can be distinguished from the zero phase shift, which will make our complete HBSA scheme more practical and realizable. In our photon number QND, the $X$-quadrature measurement is utilized, which can lead to an error probability relevant to the strength of coherent state and the phase shift. Research has shown that the error probability is less than $10^{-5}$ when $\alpha\theta^2 > 9$ \cite{Kerr1}, which indicates that our complete HBSA scheme can operate in the regime of weak cross-Kerr nonlinearity. It should be noted that the generation of Kerr nonlinearities in optical microcavities and magnonic systems have also been investigated recently \cite{Kerr14,Kerr15}. Moreover, the cavity-assisted interaction and quantum dot spin in optical microcavity may also provide accessible ways to build the photon number QND we required \cite{A9,A10,A11,A12,A13,A14,A15}.

In 2016, Li \emph {et al.} proposed the self-assisted mechanism for the complete analysis of hyperentangled state in two DOFs \cite{9}. In this paper, the self-assisted mechanism is embedded in our complete HBSA scheme for hyperentanglement in three DOFs. Our scheme can largely simplify the discrimination procedure and reduce the requirement for nonlinearity, when compared with the work of Liu \emph{et al.} \cite{12}, in which the cross-Kerr nonlinearity is also used. However, six times nonlinear quantum interactions are required in the scheme of Liu \emph{et al.}, and only three times are required in our scheme. Therefore, our scheme is easier to be realized, and will have a higher efficiency. Compared with the scheme of Yu \emph{et al.} \cite{add2}, in which the self-assisted mechanism and Kerr effect are also exploited, our scheme does not need the auxiliary entanglement in the fourth DOF. In 2007, Li \emph {et al.} investigated the distinguishability of HBSA with linear optics and auxiliary entanglement, and they found that auxiliary Bell state in the third DOF cannot accomplish the complete HBSA in two DOFs \cite{3}. Based on this theory, for the HBSA in three DOFs, two QNDs are utilized to determine the Bell states in the third DOF, and at least one more QND is required for realizing a complete three-DOF HBSA. Therefore, the three times nonlinear interactions can be viewed as the fewest times when the two-photon QND is utilized.

In summary, we have presented an efficient scheme for the complete analysis of photonic hyperentangled Bell state in three different DOFs, including the polarization DOF and two longitudinal momentum DOFs. The distinguishing process is accomplished with the help of weak cross-Kerr nonlinearity and self-assisted mechanism, which can make our scheme simple and realizable. We also have showed the applications of our complete HBSA scheme in the quantum teleportation of a single-photon state and entanglement swapping between hyperentangled pairs in three DOFs, and we believe this scheme will be useful for the future high-capacity quantum communication.

\section*{Data availability} 
Data underlying the results presented in this paper are not publicly available at this time but may be obtained from the author upon reasonable request.



\section*{References}

\begin{thebibliography}{99}
 
\bibitem{QT} Bennett C H, Brassard G, Crépeau C, Jozsa R, Peres A and Wootters W K 1993 Phys. Rev. Lett. 70 1895
\bibitem{QDC} Bennett C H and Wiesner S J 1992 Phys. Rev. Lett. 69 2881
\bibitem{ES} Zukowski M, Zeilinger A, Horne M A and Ekert A K 1993 Phys. Rev. Lett. 71 4287
\bibitem{QI} Pittman T B, Shih Y H, Strekalov D V and Sergienko A V 1995 Phys. Rev. A 52 R3429
\bibitem{add1} Kwiat P G 1997 J. Mod. Opt. 44 2173 

\bibitem{Du1} Deng F G, Ren B C and Li X H 2017 Sci. Bull. 62 46
\bibitem{Du2} Du F F and Shi Z R 2019 Opt. Express 27 17493
\bibitem{Du3} Du F F, Liu Y T, Shi Z R, et al 2019 Opt. Express 27 27046
\bibitem{Du4} Du F F, Fan G, Wu Y M and Ren B C 2023 Chin. Phys. B 32 060304 
\bibitem{Du5} Du F F, Fan G, Ren X M and Ma M 2023 Advanced Quantum Technologies 6 2300201
\bibitem{Du6} Han Y H, et al 2021 Opt. Express 29 20045
\bibitem{Du7} Xia B Y, et al 2018 Laser Phys. 28 095201 

\bibitem{A1} Kwiat P G and Weinfurter H 1998 Phys. Rev. A 58 R2623
\bibitem{A2} Walborn S P, Pádua S and Monken C H 2003 Phys. Rev. A 68 042313
\bibitem{ZZ1} Zeng Z, Wang C and Li X H 2014 Commun. Theor. Phys. 62 683
\bibitem{A3} Mattle K, Weinfurter H, Kwiat P G and Zeilinger A 1996 Phys. Rev. Lett. 76 4656
\bibitem{A4} Barreiro J T, Wei T C and Kwiat P G 2008 Nat. Phys. 4 282
\bibitem{A5} Williams B P, Sadlier R J and Humble T S 2017 Phys. Rev. Lett. 118 050501
\bibitem{A6} Hu X M, Guo Y, Liu B H, Huang Y F, Li C F and Guo G C 2018 Sci. Adv. 4 eaat9304
\bibitem{A7} Kong L J, et al 2019 Sci. Adv. 5 eaat9206
\bibitem{A8} Kong L J, et al 2019 Phys. Rev. A 100 023822
\bibitem{AD1} Wang G Y, Ren B C, Deng F G and Long G L 2019 Opt. Express 27 8994
\bibitem{AD2} Zeng Z and Zhu K D 2020 Laser Phys. Lett. 17 075203
\bibitem{AD3} Zhou X J, Liu W Q, Wei H R, Zheng Y B and  Du F F 2022 Front. Phys. 17 41502

\bibitem{2} Wei T C, Barreiro J T and Kwiat P G 2007 Phys. Rev. A 75 060305(R)
\bibitem{3} Li X H and Ghose S 2017 Phys. Rev. A 96 020303(R)
\bibitem{4} Pisenti N, Gaebler C P E and Lynn T W 2011 Phys. Rev. A 84 022340
\bibitem{5} Sheng Y B, Deng F G and Long G L 2010 Phys. Rev. A 82 032318
\bibitem{6} Ren B C, Wei H R, Hua M, Li T and Deng F G 2012 Opt. Express 20 24664
\bibitem{7} Wang T J, Lu Y and Long G L 2012 Phys. Rev. A 86 042337
\bibitem{8} Liu Q and Zhang M 2015 Phys. Rev. A 91 062321
\bibitem{9} Li X H and Ghose S 2016 Phys. Rev. A 93 022302
\bibitem{10} Li X H and Ghose S 2016 Opt. Express 24 18388
\bibitem{ZZ2} Zeng Z, Li X H, Wang C, Wang L L, Liu Z Z and Wei H 2015 Commun. Theor. Phys. 64 281
\bibitem{Z1} Zeng Z 2018 Laser Phys. Lett. 15 055204
\bibitem{11} Cao C, Zhang L, Han Y H, Yin P P, Fan L, Duan Y W and Zhang R 2020 Opt. Express 28, 2857
\bibitem{Z2} Zeng Z and Zhu K D 2020 New J. Phys. 22 083051
\bibitem{Z3} Zeng Z 2022 J. Opt. Soc. Am. B 39 2272
\bibitem{C1} Sun Y H, Guo Y Q and Cao C 2023 Quantum Inf. Process. 22 344

\bibitem{12} Liu Q, Wang G Y, Ai Q, Zhang M and Deng F G 2016 Sci. Rep. 6 22016
\bibitem{13} Wang M Y, Yan F L and Gao T 2018 Laser Phys. Lett. 15 125206
\bibitem{14} Zhang H R, Wang P, Yu C Q and Ren B C 2021 Chin. Phys. B 30 030304
\bibitem{add2} Yu C Q, Zhang Z, Qi J and Ren B C 2022 Front. Quantum. Sci. Technol. 1 985130
\bibitem{15} Zhou X J, Liu W Q, Zheng Y B, Wei H R and Du F F 2022 Ann. Phys. (Berlin) 534 2100509
\bibitem{C2} Sun Y H, Guo Y Q and Cao C 2023 J. Opt. Soc. Am. B 40 2073

\bibitem{16} Barreiro J T, Langford N K, Peters N A and Kwiat P G 2005 Phys. Rev. Lett. 95 260501
\bibitem{17} Vallone G, Ceccarelli R, De Martini F and Mataloni P 2009 Phys. Rev. A 79 030301(R)
\bibitem{add3} Wang X, Yu S, Liu S, Zhang K, Lou Y, Wang W and Jing J 2022 Advanced Photonics Nexus 1, 016002
\bibitem{add4} Achatz L, et al 2023 npj Quantum Inf. 9 45

\bibitem{Kerr1} Nemoto K and Munro W J 2004 Phys. Rev. Lett. 93 250502
\bibitem{Kerr2} Munro W J, Nemoto K and Spiller T P 2005 New J. Phys. 7 137
\bibitem{Kerr3} Shapiro J H 2006 Phys. Rev. A 73 062305
\bibitem{Kerr4} Shapiro J H and Razavi M 2007 New J. Phys. 9 16
\bibitem{Kerr5} Matsuda N, Shimizu R, Mitsumori Y, Kosaka H and Edamatsu K 2009 Nat. Photon. 3 95
\bibitem{Kerr6} Hoi I C, Kockum A F, Palomaki T, Stace T M, Fan B and Tornberg L 2013 Phys. Rev. Lett. 111 053601
\bibitem{Kerr7} Feizpour A, Hallaji M, Dmochowski G and Steinberg A M 2015 Nat. Phys. 11 905
\bibitem{Kerr8} Beck K M, Hosseini M, Duan Y H and Vuletic V 2016 PNAS 113 9740
\bibitem{Kerr9} Tiarks D, Schmidt S, Rempe G and Durr S 2016 Sci. Adv. 2 e1600036
\bibitem{Kerr10} Sinclair J, Angulo D, Lupu-Gladstein N, Bonsma-Fisher K and Steinberg A M 2019 Phys. Rev. Res. 1 033193 
    
\bibitem{Kerr11} Lin Q and He B 2009 Phys. Rev. A 80 042310
\bibitem{Kerr12} Wang C, Zhang Y and Jin G S 2011 Quantum Inf. Comput. 11 988
\bibitem{Kerr13} Cao C, Wang C, He L, et al 2013 Int. J. Theor. Phys. 52 1265
\bibitem{Kerr14} Zhang X Y, Cao C, Gao Y P, et al 2023 New J. Phys. 25 053039
\bibitem{Kerr15} Zhou Y R, Zhang Q F, Liu F F, et al 2024 Opt. Express 32 2786

\bibitem{A9} Duan L M and Kimble H J 2004 Phys. Rev. Lett. 92 127902
\bibitem{A10} Bonato C, Haupt F, Oemrawsingh S S R, Gudat J, Ding D P, van Exter M P and Bouwmeester D 2010 Phys. Rev. Lett. 104 160503
\bibitem{A11} Wang T J and Wang C 2013 J. Opt. Soc. Am. B 30 2689
\bibitem{A12} Du F F, Ren X M, Ma M and Fan G 2023 Appl. Phys. Express 16 102006
\bibitem{A13} Du F F, Wu Y M, Fan G and Ma Z M 2023 Ann. Phys. (Berlin) 535 2200507
\bibitem{A14} Cao C, Fan L, Chen X, Duan Y W, Wang T J, Zhang R and Wang C 2017 Quantum Inf. Process. 16 98
\bibitem{A15} Fan L and Cao C 2021 J. Opt. Soc. Am. B 38 1593

\end{thebibliography}
\end{document}